\documentclass[aps,prx,superscriptaddress,reprint]{revtex4-2}
\pdfoutput=1

\usepackage{silence}
\usepackage{comment}

\usepackage[utf8]{inputenc}
\usepackage[breaklinks]{hyperref}
\usepackage[british]{babel}
\usepackage[babel=true]{csquotes}

\usepackage{graphicx}
\graphicspath{{Figures/}}
\usepackage{xcolor}

\usepackage{amsmath}
\usepackage{siunitx}
\DeclareSIUnit{\fps}{ \translate{frames per second} }
\usepackage{textgreek}
\usepackage{textcomp}
\usepackage{lmodern}
\usepackage{tabularx}
\usepackage{xcolor}

\usepackage{caption}
\newcommand{\tate}{\texorpdfstring{1\textit{T$'$}-TaTe\textsubscript{2}}{1\textit{T}-TaTe2}}
\newcommand{\vte}{\texorpdfstring{1\textit{T$'$}-VTe\textsubscript{2}}{1\textit{T}-VTe2}}
\newcommand{\nbte}{\texorpdfstring{1\textit{T$'$}-NbTe\textsubscript{2}}{1\textit{T}-NbTe2}}
\newcommand{\tas}{\texorpdfstring{1\textit{T}-TaS\textsubscript{2}}{1\textit{T}-TaS2}}
\newcommand{\tase}{\texorpdfstring{1\textit{T}-TaSe\textsubscript{2}}{1\textit{T}-TaSe2}}
\newcommand{\tatese}{\texorpdfstring{1\textit{T}-TaSe\textsubscript{$2-x$}Te\textsubscript{$x$}}{1\textit{T}-TaSe2}}
\usepackage{float}
\newcommand{\uproman}[1]{\uppercase\expandafter{\romannumeral#1}}

\begin{document}

\title{Femtosecond trimer quench cycled at megahertz rates in the unconventional charge-density wave material \tate{}}

\author{Till Domröse}
\author{Claus Ropers}
\email[Corresponding author: ]{claus.ropers@mpinat.mpg.de}
\affiliation{Department of Ultrafast Dynamics, Max Planck Institute for Multidisciplinary Sciences, 37077 Göttingen, Germany}
\affiliation{4th Physical Institute -- Solids and Nanostructures, University of Göttingen, 37077 Göttingen, Germany}


\begin{abstract}
Ultrafast optical switching of materials properties is of great relevance both for future technological applications as well as gaining fundamental physical insights to microscopic couplings and nonequilibrium phenomena. Transition-metal dichalcogenides (TMDCs) combine photo-sensitivity with strong correlations, furthering rich phase diagrams and enhanced tunability. Owing to its chemical composition, \tate{} exhibits an electronically and structurally unique set of charge-density waves (CDWs), featuring increased conductivity and a reduced prominence of amplitude modes. Compared to other charge-ordered TMDCs, only very few studies addressed the ultrafast response of this material to optical excitation. In particular, the question whether such unconventional properties translate to unusual quench dynamics remains largely unresolved. Here, we investigate the structural dynamics in \tate{} by means of ultrafast nanobeam electron diffraction. The experiments are carried out with a tailored sample design that allows for excitation at \SI{2}{\MHz} repetition rate, higher than any structural phase transformation probed thus far. Harnessing the enhanced resolution and sensitivity of this approach, we reveal a strongly directional cooperative atomic motion during the one-dimensional quench of the low-temperature trimer lattice. These dynamics are completed within less than \SI{500}{\fs}, substantially faster than reported previously. In striking contrast, the periodic lattice distortion of the room-temperature phase is unusually robust against high-density electronic excitation. In conjunction with the known sensitivity of \tate{} to chemical doping, we thus expect the material to serve as a versatile platform for tunable structural control by optical stimuli.
\end{abstract}

\maketitle

\section{Introduction}
Tailoring electronic properties by chemical doping is a central element of current semiconductor technology. Beyond established applications, transition metal dichalcogenides (TMDCs), strongly correlated materials in a chalcogen-metal-chalcogen trilayer structure \cite{Wilson1975}, promise novel functionality \cite{Jariwala2014}, as comparably small modifications of the electronic state can evoke large effects via couplings to phononic, orbital or spin degrees of freedom \cite{Tokura2017,basov2017}. In these compounds, the quasi-two-dimensional characteristics in conjunction with strong electron-phonon interactions favour the occurrence of various correlated phases \cite{Li2021}, including charge density waves (CDWs) coupled to a periodic lattice distortion (PLD) \cite{Rossnagel2011}. The corresponding low-energy band gap is highly susceptible to external stimuli \cite{Yoshida2015,Ugeda2016,schmitt2022,jarc2023} or subtle alterations of the chemical composition \cite{ang2013,luo2016} and atomic configurations \cite{sung2023}.

Alongside structural tunability, TMDCs have proven interesting targets for the functionalization of non-equilibrium photodoping \cite{basov2017,delatorre2021,bao2022} and in optoelectronics \cite{mak2016}. Probing the interplay of structural and electronic excitations \cite{siders1999,lindenberg2000,sokolowski-tinten2003,siwick2003,beaud2014,gerber2017,Waldecker2017,Wall2018,Sie2019,tauchert2022,filippetto2022,buzzi2018,alcorn2023,Storeck2020,Otto2021,Cremons2017,mann2016,Perfetti2008,Rohwer2011,Eichberger2010,erasmus2012,Sohrt2014,Han2015,Haupt2016,LeGuyader2017,Zong2018,sun2018,Laulhe2017,Vogelgesang2018,domrose2023,ji2020,Danz2021b,chen2016,cheng2022,ji2020,zhang2022,zong2019a,horstmann2020,Kogar2020,singer2016,Stojchevska2014,gerasimenko2019a}, ultrafast measurement schemes frequently uncover underlying fundamental processes. In the past, pulsed laser excitation has not only been used to drive transitions between different CDW phases \cite{Eichberger2010,erasmus2012,Sohrt2014,Han2015,Haupt2016,LeGuyader2017,sun2018,Laulhe2017,Zong2018,Vogelgesang2018,zong2019a,horstmann2020,ji2020,Danz2021b,chen2016,cheng2022,domrose2023,zhang2022}, but also to transiently induce \cite{Kogar2020} or enhance CDW order \cite{singer2016} as well as distinct types of disorder \cite{Laulhe2017,Vogelgesang2018,zong2019a,domrose2023}, prompt dimensional crossovers \cite{chen2016,cheng2022,domrose2023}, and to drive materials into thermodynamically inaccessible metastable states \cite{Stojchevska2014,gerasimenko2019a,maklar2023}.

To date, the investigation of ultrafast CDW dynamics and phase transitions in TMDCs largely focused on sulphur- and selenium-based compounds \cite{Storeck2020,Otto2021,Cremons2017,mann2016,Perfetti2008,Rohwer2011,Eichberger2010,erasmus2012,Sohrt2014,Han2015,Haupt2016,LeGuyader2017,Zong2018,sun2018,Laulhe2017,Vogelgesang2018,ji2020,Danz2021b,domrose2023,chen2016,cheng2022,ji2020,zhang2022}. In contrast, the CDW phases in TMD tellurides have only recently become the subject of ultrafast studies \cite{nakamura2020,siddiqui2021,hu2022,tuniz2023,suzuki2023a}. In these materials, an enhanced hybridization of the tellurium \textit{p-} with the metal \textit{d-}bands decisively alters the transition metal valence and thus the electronic texture~\cite{canadell1992,sharma2002}. Among other features, this results in large PLD amplitudes in $(3\times1)$ superstructures shared by \vte{}, \nbte{} and \tate{}, which are stable at room temperature and above \cite{bronsema1984,sorgel2006,katayama2023}.

\begin{figure*}[t]
\centering
\includegraphics[width=\textwidth]{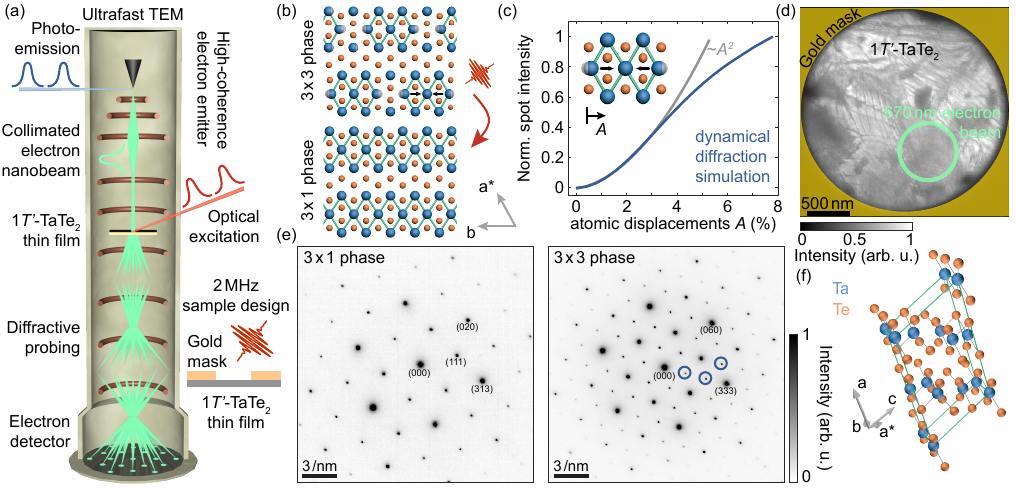}
\caption{
Nanobeam electron diffraction at megahertz rates in an ultrafast transmission electron microscope (UTEM).
(a) Schematics of the UTEM. Ultrashort electron pulses (green) are generated via linear photoemission (blue) from a tip-shaped emitter. The exceptionally high transverse beam coherence allows for collimated diffractive probing with nanometric electron illumination. Optical excitation at variable photon energy (red) drives reversible dynamics in the \tate{} thin film.
(b) In-plane crystal structure of the low-temperature $(3\times3)$ (top) and the room-temperature $(3\times1)$ phases of \tate{} (bottom). The phase transformation leads to an additional tripling of the unit cell along the $b$-axis and the formation of heptamers.
(c) Dynamical diffraction simulations yield the quantitative PLD amplitude $A$ of the $(3\times3)$ phase as a function of the spot intensity (specified in units of the interatomic distances in the $(3\times1)$ phase). For small atomic displacements, the intensity of $(3\times3)$ diffraction spots (blue curve) increases quadratically with $A$ (grey).
(d) The \tate{} thin film is suspended below a circular gold aperture to enable reversible megahertz driving of the phase transformation. In the diffraction experiments, we probe a sample area of enhanced spatial homogeneity with a diameter of \SI{670}{\nm} (green circle).
(e) Electron diffractograms of both phases recorded in the $[\bar{1}01]$ zone axis. In the $(3\times3)$ phase, the heptamer formation brings about PLD diffraction spots (highlighted by blue circles) in-between the bright reflections of the 1\textit{T} symmetry present in both phases.
(f) Monoclinic unit cell (green) of \tate{} in the $(3\times1)$ phase. Trilayers of tantalum atoms (blue) sandwiched between tellurium atoms (orange) are stacked on top of each other, with a threefold stacking sequence in both phases.
}
\label{fig:1}
\end{figure*}
 
This chemical perspective on CDW formation and the corresponding emergence of tightly-bound zigzag chains \cite{canadell1992} particularly applies to \tate{} \cite{gao2018,petkov2020,katayama2023}, the compound with the most pronounced structural distortions. In this material, the trimerization even persists up to highest temperatures, i.e., an undistorted phase is absent from the phase diagram. Upon cooling, a second transition brings about a unique $(3\times3)$ phase \cite{sorgel2006,feng2016,chen2018,elbaggari2020} that is in competition with other structural instabilities at surfaces and in few-layer systems \cite{kar2021,hwang2022,dibernardo2023}. Remarkably, the formation of the $(3\times3)$ phase results in a decrease of the electrical resistivity \cite{chen2017}, as opposed to related transitions in other TMDCs where the opening of the CDW band gap promotes the conventional insulating behavior~\cite{Wilson1975,mulazzi2010,Yoshida2015,sun2018,sun2020,meng2020}.

Its unique phase diagram and the associated unusual macroscopic properties call for further investigations of the CDW phases in \tate{}. In equilibrium, the origin of the low-temperature butterfly-like structure has been linked to an anisotropic phononic softening \cite{sorgel2006,katayama2023}, driven by Fermi-surface nesting \cite{luo2021,lin2022}. Few ultrafast experiments studied non-equilibrium dynamics in this material, finding an only weak amplitude mode \cite{hu2022} and a considerably slower partial PLD suppression \cite{siddiqui2021} compared to dynamics in other strongly-correlated materials \cite{Eichberger2010,erasmus2012,Sohrt2014,Han2015,Haupt2016,LeGuyader2017,sun2018,Laulhe2017,Zong2018,Vogelgesang2018,zong2019a,horstmann2020,ji2020,Danz2021b,chen2016,cheng2022,domrose2023,zhang2022,Hellmann2012b,tuniz2023,suzuki2023a,yusupov2010}. However, possible microscopic origins for a slower response remain to be explored, and a complete quench of the $(3\times3)$ phase has yet to be observed. Moreover, also the sensitivity of the $(3\times1)$ periodicities to photodoping both at low and at high temperatures is of considerable interest.

\begin{figure}[htbp]
\centering
\includegraphics[width=\columnwidth]{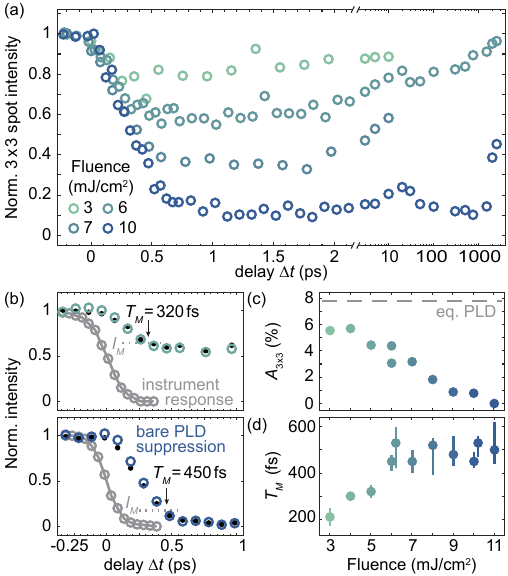}
\caption{
Femtosecond quenching of periodic lattice distortions in \tate{}.
(a) Representative delay curves derived from the intensities of second-order $(3\times3)$ diffraction spots. After a pronounced initial suppression, the PLD partially recovers for low and intermediate fluences. The material retains its initial configuration after \SI{3}{\ns}.
(b) PLD dynamics around time zero for an excitation fluence of \SI{6}{\mJ\per\cm^2} (top) and \SI{10}{\mJ\per\cm^2} (bottom), respectively. The measured curves (black circles) display a clear temporal separation from the independently recorded instrument response function (grey). As a result, the deconvoluted suppression of the $(3\times3)$ PLD (colored circles) is well resolved.
(c) Initial suppression of the PLD amplitude $A_{3\times3}$ from an equilibrium value of \SI{7.8}{\percent} for different pump fluences, based on the quantitative relation between diffraction spot intensities and atomic displacements displayed in Fig.~\ref{fig:1}(c). A complete phase transition occurs for fluences above \SI{8}{\mJ\per\cm^2}.
(d) PLD melting times $T_M$ derived from the deconvoluted delay curves displayed in (b). The phase transition slows down upon approaching the threshold excitation, indicative of a softening of the responsible structural mode.
}
\label{fig:2}
\end{figure}

In this work, we address these open questions in an investigation of the laser-induced dynamics in \tate{} with high resolution and high sensitivity. Specifically, we elucidate the transformation from the $(3\times3)$ to the $(3\times1)$ phase by means of ultrafast nanobeam electron diffraction (nano-UED). Using an unprecedented repetition rate of \SI{2}{\MHz}, we find that the structural dynamics unfold more than ten times faster than previously reported. Furthermore, the timescale of the structural quench is a strong function of excitation fluence, with a transient suppression of the lattice distortion completed as fast as \SI{200}{\fs}. Approaching the threshold excitation, the phase transition proceeds within \SI{500}{\fs}, indicative of a softening of the corresponding structural mode. In contrast, we identify an atypical resilience of the high-temperature $(3\times1)$ superstructure to the optical excitation, revealing the coexistence of seemingly similar types of structural order with completely different origins suspended within the same host material.

\section{Experiment}

In the experiments, we excite a \tate{} thin-film in the low-temperature phase with ultrashort laser pulses (\SI{800}{\nm} wavelength, \SI{50}{\fs} duration, between \SI{0.7}{\mJ\per\square\cm} and \SI{11}{\mJ\per\square\cm} fluence), and probe the subsequent dynamics with collimated ultrashort electron pulses (\SI{120}{\keV} energy, \SI{200}{\fs} duration, \SI{670}{\nm} beam diameter). To this end, the Göttingen Ultrafast Transmission Electron Microscope [UTEM, Fig.~\ref{fig:1}(a), see also Methods] is equipped with a tip-shaped emitter whose nanometric size results in an exceptionally high transverse beam coherence \cite{Feist2017}. The small effective source size promotes picometer beam emittances \cite{Feist2017} and high reciprocal-space resolution even for particularly narrow, collimated electron beams with nanometer diameters \cite{domrose2023}. The instrument thus enables high-resolution ultrafast electron diffraction without spatial averaging over microscopic sample heterogeneity. Importantly, a tailored sample design [c.f. Fig.~\ref{fig:1}(d)] facilitates optimized thermal dissipation \cite{Danz2021b,domrose2023}, which in turn allows us to study the reversible dynamics at a repetition rate of \SI{2}{\MHz}, higher than in any previous measurement of a structural phase transition. In this way, our measurements drastically enhance the sensitivity to low-intensity features and eliminate common issues in the stroboscopic investigation of thin material films \cite{kazenwadel2023}.

\subsection{CDW phases in \tate{}}

The basic monoclinic structure of \tate{} is a commensurate distortion of the 1\textit{T} symmetry commonly found in TMDCs. This room-temperature $(3\times1)$ phase exhibits a tripling of the (hypothetical) undistorted unit cell along the lattice vector $\textbf{a*}$ and a three-fold layer stacking sequence [Fig.~\ref{fig:1}(f)] \cite{sorgel2006,elbaggari2020}. This formation of zigzag chains predominantly results in atomic displacements in a direction perpendicular to the lattice vector $\textbf{b}$ [Fig.~\ref{fig:1}(b)] with an amplitude that amounts to up to \SI{12.4}{\percent} of the average interatomic distance. Below a temperature of \SI{174}{\kelvin}, a further \SI{7.8}{\percent} distortion, mainly oriented along the $\textbf{b}$-direction, transforms the crystal into the $(3\times3)$ phase \cite{sorgel2006,feng2016,chen2018}. Both phases, as well as the transition between them, primarily involve re-configurations of the tantalum sublattice \cite{sorgel2006} and band structure, suggesting an orbital-selective mechanism connected to the PLD formation \cite{mitsuishi2023}.

\subsection{Femtosecond structural melting}

\begin{figure*}[t]
\centering
\includegraphics[width=\textwidth]{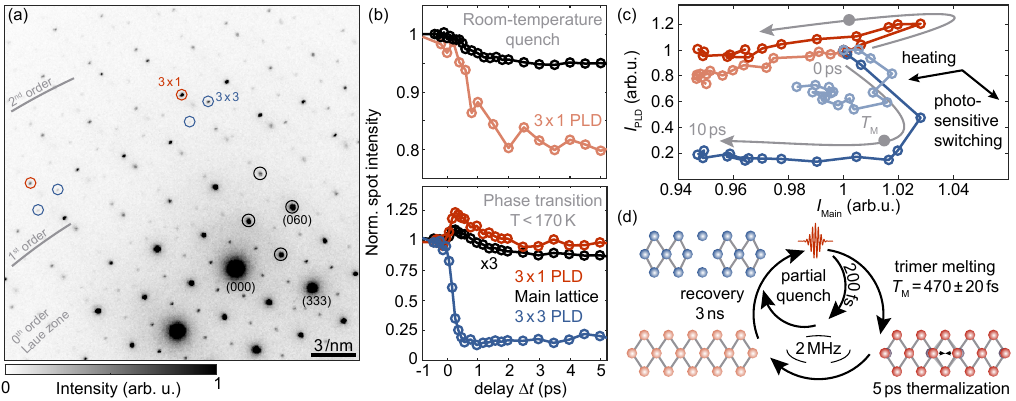}
\caption{
Ultrafast dynamics in the first-order Laue zone and phase transition mechanism.
(a) Example electron diffractogram for an electron beam tilted out of the [$\bar{1}01$] zone axis. The individual signatures of the $(3\times1)$ (red circles) and the $(3\times3)$ structure (blue) reside in the first-order Laue zone, whereas the zero-order Laue zone contains second-order PLD diffraction spots and those of the undistorted 1\textit{T} lattice (black).
(b) Temporal evolution of first-order $(3\times3)$ (blue), $(3\times1)$ (red) and of the undistorted structural signatures (black, \SI{10}{\mJ\per\cm^2} fluence). At room temperature (top), we solely observe the few-picosecond Debye-Waller effect. During the phase transition (bottom), the $(3\times3)$ spots behave similarly to the second-order PLD spots, whereas the remaining spots display a transient overshoot before a slower, few-percent suppression.
(c) PLD spot intensities as a function of the main lattice spot intensities, visualizing the two dynamical contributions to the spot evolution (black arrows). As the main lattice reflections only display thermal dynamics, a coinciding temporal behaviour of the PLD demonstrates the low susceptibility of the high-temperature periodicities to the optical stimulus both below (dark red) and above the phase transition temperature (light red). In contrast, the $(3\times3)$ PLD can efficiently be quenched (light blue, \SI{5}{\mJ\per\cm^2} fluence) or switched (dark blue, \SI{10}{\mJ\per\cm^2}) by laser excitation.
(d) Phase transition mechanism resulting from deciphering the individual structural Fourier components in (c). The optical excitation launches one-dimensional atomic motions leading to sub-picosecond trimer melting towards the high-temperature $(3\times1)$ distortion, followed by a \SI{5}{\ps} thermalization. The system recovers its equilibrium low-temperature configuration after only \SI{3}{\ns}. For a partial quench, the initial trimer suppression unfolds as fast as \SI{200}{\fs}.
}
\label{fig:3}
\end{figure*}

In a first measurement series, we characterise the structural dynamics upon laser excitation of the $(3\times3)$ phase at different fluences. We set the electron beam to perpendicular incidence onto the material layers spanned by the vectors $\textbf{b}$ and $\textbf{a*}$, probing reciprocal space in the [$\bar{1}01$] zone axis [see Fig.~\ref{fig:1}(e) for the corresponding electron diffractograms of both phases]. In these images, diffraction spots $(hkl)$ with $k=3n$ (where $n$ is an integer) correspond to the undistorted structure [hereafter referred to as main lattice spots, see Appendix \ref{sec:RecLattice} for a more detailed description of the reciprocal lattice]. In contrast, the $(3\times3)$ distortion produces the weaker reflections in-between, analogously to the second-order CDW spots observed in the distorted phases of related 1\textit{T} polytype TMDCs \cite{Wilson1975}. Dynamical diffraction simulations supply a quantitative relationship between measured intensities and the PLD amplitude, thus presenting a precise measure of the order parameter for the $(3\times3)$ to $(3\times1)$ transition [c.f. Fig.~\ref{fig:1}(c), and Appendix \ref{sec:fluencescans}].

Tracking the intensities of the $(3\times3)$ satellites at variable temporal delay $\Delta t$ between the laser pump and the electron probe pulses yields the curves shown in Fig.~\ref{fig:2}(a) [see also Appendix \ref{sec:fluencescans} and Fig.~\ref{fig:FluenceScan}]. Therein, the observed temporal characteristics of the optically-induced structural dynamics spans four orders of magnitude. Following a rapid femtosecond suppression of the PLD amplitude, we observe a partial recovery after \SI{5}{\ps} [Fig.~\ref{fig:2}(b)] for low and intermediate pump fluences. For fluences above \SI{8}{\mJ\per\cm^2}, on the other hand, the $(3\times3)$ intensities remain suppressed, evidencing the transition into the $(3\times1)$ phase. From here, the initial state is re-established after only \SI{3}{\ns}, leaving sufficient time for further relaxation and heat dissipation before the arrival of the subsequent pump pulse after \SI{0.5}{\us}. Overall, we find no significant dependence of the dynamics on the excitation wavelength in the visible spectral range [c.f. Fig.~\ref{fig:WavelengthScan}].

A critical parameter retrieved from such delay curves is the switching time of the structural transition as it is linked to the degrees of freedom involved in the phase formation, thereby elucidating microscopic origins of the symmetry-broken state also in thermal equilibrium \cite{Hellmann2012b}. Femtosecond features of the dynamics frequently approach the temporal resolution of ultrafast electron diffraction experiments given by the electron pulse length. Typically, both the exact temporal overlap as well as the pulse shape and duration are indirectly inferred from reference measurements on separate samples. Here, we provide a quantitative approach to the determination of the ultrafast instrument response function, which we believe will serve as highly valuable benchmark for future ultrafast imaging and diffraction studies. Specifically, harnessing the versatility of the UTEM, we spectroscopically determine the inelastic scattering of electrons by the pump-induced optical near-field at the sample surface \cite{barwick2009,Feist2015a,dahan2021} [see Appendix \ref{sec:electronpulses} and Fig.~\ref{fig:Pulse duration}]. This instantaneous field-driven interaction not only allows us to determine time zero, but also yields a precise measure of the electron pulse shape. In particular, we record such reference measurements alongside every delay curve, showing consistent electron pulse durations of about \SI{200}{\fs}. The corresponding instrument response functions can be used to exemplify how an instantaneous signal drop would be probed [grey curves in Fig.~\ref{fig:2}(b)]. Its clear temporal separation from the measured delay curves [black points in Fig.~\ref{fig:2}(b)] demonstrates that the laser-induced phase switch is quantitatively resolved. For completeness, measuring the precise electron pulse shape allows for a deconvolution of the recorded signal by the instrument response [colored circles], which, as expected, yields only a small correction.

Generally, the temporal suppression of the $(3\times3)$ PLD amplitude evolves in a manner expected for a displacive excitation of the nuclei after a sudden quench of the electronic subsystem \cite{zeiger1992}. However, a fit of the corresponding cosinusoidal function slightly underestimates the changes to the PLD amplitude in the first few hundred femtoseconds of the dynamics, both for the measured as well as the deconvoluted delay curves [see Fig.~\ref{fig:FluenceScan}]. This observation hints at either anharmonicities in the displacive potential or a (partial) contribution from an impulsive mechanism \cite{merlin1997} during the initial stage of the dynamics. Consequently, we quantify the melting times directly in terms of the measured intensity evolution, evaluating the time delay for a relative suppression of \SI{85}{\percent} with respect to the minimum value for each fluence. We find PLD quenching times as small as \SI{200}{\fs}, more than ten times faster than previously reported (\SI{1.4}{\ps} decay time to $1/e$ intensity level in ref.~ \cite{siddiqui2021}). These results underline the strongly-correlated nature of the $(3\times3)$ phase, which, accordingly, can be rapidly quenched via an optically-induced electronic transition. Furthermore, the quench time increases upon approaching the phase transition threshold, reaching a plateau at \SI{470\pm20}{\fs} for a complete phase switch [Fig.~\ref{fig:2}(d)]. Previously, a slowing-down of structural dynamics in CDW systems on sub-picosecond timescales was linked to a softening of the corresponding structural mode \cite{yusupov2010}, and also identified in all-optical measurements of \tate{} \cite{hu2022}. Indeed, a comparatively flat potential landscape close to the critical point of the transition is expected to minimize the forces driving the required structural modifications, particularly for second-order and weakly first-order transitions \cite{zong2019}. Moreover, faster dynamics could result from spatial carrier dynamics within the depth of the surface-excited film, while above-threshold excitation was shown to switch the PLD completely also for a comparably thick \tas{} film \cite{Danz2021b}. Given the significant discrepancy between our findings and those of Ref. \cite{siddiqui2021}, some comments on a possible explanation are warranted. As a possible source of a delayed response in probing a mesoscopic sample, spatiotemporal carrier dynamics resulting from heterogeneous excitation, e.g. at edges, may be a contributing factor. In contrast, as our wavelength-dependent data indicate, differences in optical excitation are likely not responsible for these dissimilar timescales.

\subsection{Phase transformation mechanism}

Beyond this precise temporal characterization of the phase transition, the stability of the high-temperature phase in thermal equilibrium naturally questions the photodoping sensitivity of the $(3\times1)$ periodicity that is present in both the low- as well as the high-temperature structure. In a second measurement series, we therefore more closely investigate the microscopic processes during the dynamics, accessing the different Fourier components of the $(3\times1)$ and the $(3\times3)$ modifications individually. To this end, we change the relative angle of incidence between the electron beam and the sample surface by approximately \SI{0.7}{\degree}, probing the more intense 'first-order' superstructure spots. Specifically, these spots are situated in the first-order Laue zone (FOLZ) and at one third of the distance between neighbouring main lattice reflections, owing to the tripling of the unit cell both within as well as perpendicular to the individual material layers [Fig.~\ref{fig:3}(a)]. 

After laser excitation, the $(3\times3)$ spots are suppressed within \SI{500}{\fs}, similar to their second-order counterparts, followed by a few-picosecond recovery for intermediate pump fluences [cf. Fig.~\ref{fig:3}(b) and Fig.~\ref{fig:First-order PLD}]. In contrast, the delay curves of the $(3\times1)$ lattice periodicities [red curves in Fig.~\ref{fig:3}(b)] display the same timescales as for the low-temperature periodicities, but otherwise opposite behaviour. The early-stage dynamics are characterised by an intensity increase of up to \SI{20}{\percent} for the highest pump fluence, followed by a slower, few-picosecond suppression. These fastest features, however, are only an indirect consequence of the $(3\times3)$ melting, as we find a similar behaviour for the main lattice spots [black curves in Fig.~\ref{fig:3}(b)]. During the phase transformation, the relative weight of both of these periodicities in the Fourier representation of the crystal lattice is enhanced, such that diffracted intensity is re-distributed from the low- to the high-temperature reflections, leading to the observed intensity overshoot.

Based on these observations, we deduce that the phase transformation proceeds as illustrated in Fig.~\ref{fig:3}(d). The optical excitation induces a quench specifically of the trimerization oriented along the lattice vector \textbf{b}, i.e., of the $(3\times3)$ type. The atomic displacements along the perpendicular direction, on the other hand, remain largely unaffected and persist even in the out-of-equilibrium state shortly after time zero. The presumed distinction between a soft $(1\times3)$ and a stable $(3\times1)$ crystal axis after every pump-probe cycle becomes apparent by comparing the PLD delay curves to those of the "photoinert" main lattice symmetries [black curves in Fig.~\ref{fig:3}(b)]. A joint temporal evolution along a close-to-horizontal trajectory in the representation displayed in Fig.~\ref{fig:3}(c) signifies a phonon-induced peak suppression, indicative solely of an overall rise in lattice temperature, i.e., the Debye-Waller effect \cite{Waldecker2017,Otto2021}. Variations towards the lower-right corner of the diagram, on the other hand, is evidence of a strong photo-sensitivity and the associated trimer quench. Clearly, the high-temperature distortion can be assigned to the former case, while the low-temperature PLD is of the latter type. This remarkable difference has immediate consequences also for the non-equilibrium pathway between both phases. Specifically, we observe no sign of transient disorder during the phase transformation, while also the out-of-plane periodicity is unaffected by the optical excitation [cf. Fig.~S4]. As a result, the phase switching unfolds completely reversibly also after a full suppression of the low-temperature order, resulting in a unidirectional transition between a three- and a two-dimensional superstructure, induced by the \SI{2}{\MHz} drive.

\section{Discussion}

The unusually large PLD amplitudes of up to \SI{12.4}{\percent} in the room-temperature phase and \SI{7.8}{\percent} in the low-temperature phase of \tate{} suggest an almost molecular description of the crystal structure, possibly outside the typical description of CDW phases \cite{canadell1992,sharma2002,gao2018,petkov2020,katayama2023}. While the photosensitivity of the $(3\times3)$ distortions is the same as found for the CDW phases of other TMDCs, the remarkable stability of the $(3\times1)$ phase, both with respect to the excitation of the low-temperature PLD as well as the synonymous absence of the 1\textit{T} polytype in thermal equilibrium, supports this picture. The difference between the two superperiodicities in this regard even extends to highly non-equilibrium scenarios, as apparent from a reference measurement where we directly excite the $(3\times1)$ phase at room temperature. Instead of a femtosecond CDW quench corroborated by a transient intensity overshoot of the main lattice reflections, analogously to the phase transition at low temperatures, we again solely find a thermalisation of the entire lattice also for comparably high excitation densities of \SI{10}{\mJ\per\cm^2} [Fig.~\ref{fig:3}(b) and (c)]. In other words, the $(3\times1)$ PLD is unusually stable against photodoping throughout the entire phase diagram of the material, and not even a sudden increase of the electronic temperature to several thousand Kelvin as typically achieved by pulsed laser excitation \cite{dolgirev2020a} suffices to transiently suppress the $(3\times1)$ order.

Our results underline that the two superstructures supported by the material originate from completely different driving forces, i.e., a perturbational and thus highly tunable distortion interwoven with a robust and only superficially related lattice trimerization. Thus, a femtosecond quench of the $(3\times1)$ phase is either not achievable, or might require an orbital-selective excitation at different photon energies \cite{mitsuishi2023,siddiqui2021}. This unique characteristic among the CDW phases in TMDCs opens up interesting perspectives on exploring the origins of CDW formation in a broader sense. In particular, the $(3\times3)$ phase appears to be the structural analogue of the $(3\times1)$ in the other group-\uproman{5} compounds in terms of PLD amplitude, temperature-dependence and photosensitivity \cite{bronsema1984,zhang2022a,tuniz2023,suzuki2023a}. One may therefore speculate whether the additional $(3\times1)$ distortion in \tate{} may serve as a platform to tune the optically-induced behaviour of the material by changing the orbital texture, partially substituting the tellurium atoms with selenium in \tatese{}. In thermal equilibrium, the $(3\times3)$ phase was shown to occur for doping levels down to $x>1.3$, followed by closely related but incommensurate distortions for lower tellurium contents \cite{wei2017}.

\section{Conclusion}

Ultrafast pencil-beam electron diffraction enables a precise mapping of the laser-induced dynamics in \tate{} free from the influence of structural heterogeneity. Our results suggest a close correspondence between the driving mechanisms responsible for the CDW formation at low temperatures to other related CDW phases. At the same time, the characteristics of the high-temperature phase further exemplifies the unique position of the material within the family of TMDCs, warranting further investigations in the ultrafast time domain by, e.g., selectively exciting specific electronic transitions. Furthermore, the peculiar phase diagram render \tate{} a promising component in heterostructures combining different layered materials \cite{Geim2013,hoque2020}. The pronounced contrast enhancement and the possibility to further minimize the electron probe beam diameter offered by megahertz driving will extend the experimental possibilities to investigate non-equilibrium dynamics in such systems. Finally, we anticipate that the precise determination to the electron pulse shape in conjunction with the high temporal resolution of nano-UED will advance the quantitative characterization of ultrafast structural dynamics and the underlying fundamental processes for a wide range of materials and heterostructures.

\begin{acknowledgments}
The authors thank L.C. da Camara Silva for technical support in the specimen preparation, and M.~Sivis for technical support in focused ion beam milling. Furthermore, we gratefully acknowledge insightful discussions with S.~Mathias, J.H.~Gaida, F.~Kurtz, M.~Franz and I.~Vinograd, as well as continued support from the Göttingen UTEM team.
This work was funded by the Deutsche Forschungsgemeinschaft (DFG, German Research Foundation) in the Collaborative Research Centre ``Atomic scale control of energy conversion'' (217133147/SFB 1073, project A05) and via resources from the Gottfried Wilhelm Leibniz Prize (RO 3936/4-1).
\end{acknowledgments}





\appendix
\section{Experimental details}

\subsection{Ultrafast Transmission Electron Microscopy}
\label{sec:UTEM}
The Göttingen UTEM is based on a \enquote{JEOL JEM-2100F}, equipped with a ZrO/W Schottky emitter \cite{Feist2017}. Access to the ultrafast time domain is enabled by operating the electron source below the threshold for continuous electron emission, gating the linear photoemission of femtosecond electron pulses with pulsed laser illumination incident on the emitter tip (\enquote{Light Conversion CARBIDE}, \SI{40}{\fs} pulse duration, \SI{515}{\nm} wavelength after frequency doubling, \SI{2}{\MHz} repetition rate). For the optical excitation of the sample, a part of the laser output is fed into an optical parametric amplifier (\enquote{Light Conversion ORPHEUS F}), and converted to a wavelength between \SI{690}{\nm} and \SI{940}{\nm}.

The data is recorded by varying the relative timing between the laser pump and the electron probe pulses in the sample plane, integrating over the average response of the sample to the optical stimulus for \SI{90}{\s} at every temporal delay following the stroboscopic principle. The resulting time-resolved diffractograms and the real-space images of the specimen were recorded with a direct electron detector (\enquote{Direct Electron DE-16}), and processed by an electron counting algorithm, while static diffractograms were taken with a CCD camera (\enquote{Gatan Orius}). For the pulse length measurements, we used a hybrid pixel detector based on the Timepix chip architecture (\enquote{Amsterdam Scientific Instrument EMCheeta}), attached behind an electron spectrometer and energy filtering device (\enquote{CEOS CEFID}).

\subsection{Sample preparation}
\label{sec:sample}

The specimen was prepared by Ultramicrotomy from a commercially available \tate{} single crystal (\enquote{HQ graphene}). \tate{} thin-films with a nominal thickness of \SI{50}{\nm} were transferred onto a silicon nitride membrane prepared for high-repetition rate driving by covering the backside of the sample carrier with an opaque gold film (thickness \SI{250}{\nm}) via sputtering. Electron-transparency within a \SI{2}{\um} field-of-view was ensured by placing the \tate{} thin-film atop a hole drilled by ion beam milling, resulting in the sample geometry displayed in Fig.~\ref{fig:1}.

\section{Indexing of diffraction patterns}
\label{sec:RecLattice}

\begin{figure}[b]
\centering
\includegraphics[width=\columnwidth]{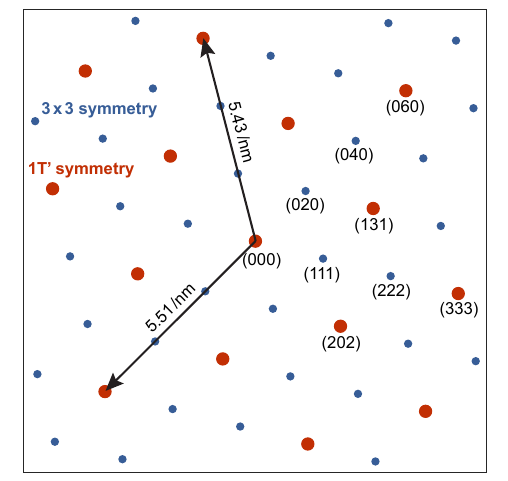}
\caption{
Reciprocal lattice of the $(3\times3)$ phase in the projection along the [$\bar{1}01$] zone axis in the zero-order Laue zone as recorded in the experimental diffractograms. The monoclinic crystal system introduces non-trivial spot-indexing of main lattice 1\textit{T'} (red) and superstructure reflections (blue) compared to the conventional nomenclature of the 1\textit{T} polytype. Note the different distances of second-order main lattice spots from the origin indicated by the black arrows.
}
\label{fig:RecLattice}
\end{figure}

As described in the main text, the periodic lattice distortions in \tate{} are commensurate with the underlying, hypothetical 1\textit{T} structure. As such, it is possible to describe the entire three-dimensional crystal in all phases with only three lattice vectors as a monoclinic crystal system, as opposed to the incommensurate phases found in other TMDC's (where the PLD superlattice is usually accounted for by two additional lattice vectors). In our experiments, we record diffractograms under perpendicular incidence with respect to the individual material layers. In these images, the brightest reflections in a close to hexagonal arrangement are a signature of the reminiscent 1\textit{T} polytype when illuminated in the [$001$] zone axis. For the choice of a monoclinic unit cell, however, this setting corresponds to the [$\bar{1}01$] zone axis, and the diffractograms require a different indexing of the depicted lattice planes, while considering systematic absences associated with the space group of \tate{}, that is, $C2/m$ (spacegroup 12, cell choice 1 in ref.~\cite{tables}). Specifically, the [$\bar{1}01$] zone axis generally only features those spots $(hkl)$ where $h=l$, while, additionally, only reflections with $h+k=2n$, $h=2n$ for $(h0l)$, $k=2n$ for $(0kl)$, $h+k=2n$ for $(hk0)$, $k=2n$ for $(0k0)$, and $h=2n$ for $(h00)$ are symmetry-allowed (where $n$ is an integer). Fig.~\ref{fig:RecLattice}(a) shows the resulting positions of reciprocal lattice points in the zero-order Laue zone, as recorded in the experiments for the $(3\times3)$ phase, and the corresponding indices $(hkl)$. Reflections with $k=3n$ describe the undistorted 1\textit{T} periodicities, and, in this projection of the reciprocal lattice and distinct from the 1\textit{T} polytype, the hexagons comprised out of spots with comparable distances to the direct beam are slightly distorted. Reflections in-between correspond to the $(3\times3)$ superperiodicity, analogously to the second-order CDW spots in the CDW phases of, e.g., \tas{} and \tase{} \cite{Wilson1975}. First-order reflections of both phases appear in the first-order Laue zone, i.e., either at larger wave vectors or upon tilting the sample. For the $(3\times3)$ unit cell, their indexing fulfills $h=l+1$ in addition to $k=3n$ for the $(3\times1)$ and $k=3n\pm1$ for the $(3\times3)$ spots.

\section{Experimental delay curves and phase switching times}
\label{sec:fluencescans}

\begin{figure*}[htbp]
\centering
\includegraphics[width=\textwidth]{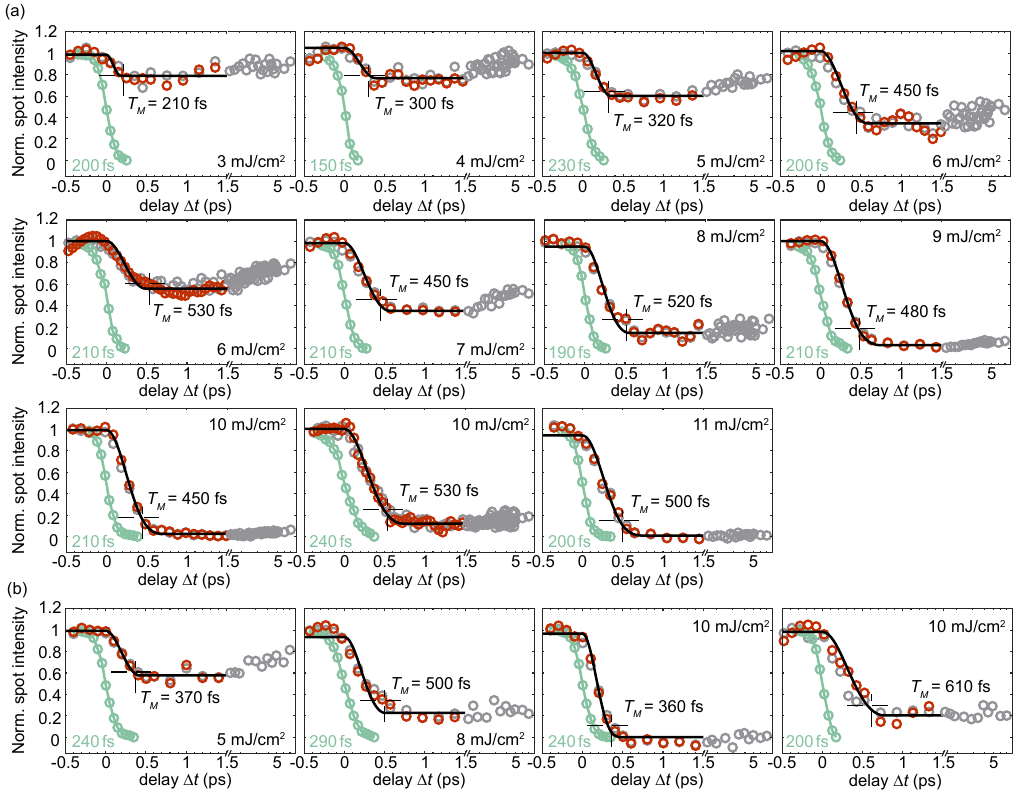}
\caption{
Fluence-dependent delay curves of second order CDW spots.
(a) Full set of delay curves recorded in the [$\bar{1}01$] zone axis for all pump fluence as discussed in Fig.~2 and in the main text, derived from the temporal evolution of second order PLD $(3\times3)$ spots. For a precise temporal evaluation of the phase transition, the measured values (grey circles) are deconvoluted (red circles) by the respective instrument response function (green) recorded for every delay scan. The specified temporal resolution corresponds to the FWHM of the respective electron pulse shape. While the melting time $T_M$ is estimated from an \SI{85}{\percent} drop of the initial value with respect to the maximum suppression (black crosses), red lines represent the cosinusoidal fit for reference.
(b) Second-order PLD spot intensities during the dynamics similar to (a), but for a tilted illumination as displayed in Fig.~3 in the main text.
}
\label{fig:FluenceScan}
\end{figure*}

\begin{figure}[htbp]
\centering
\includegraphics[width=\columnwidth]{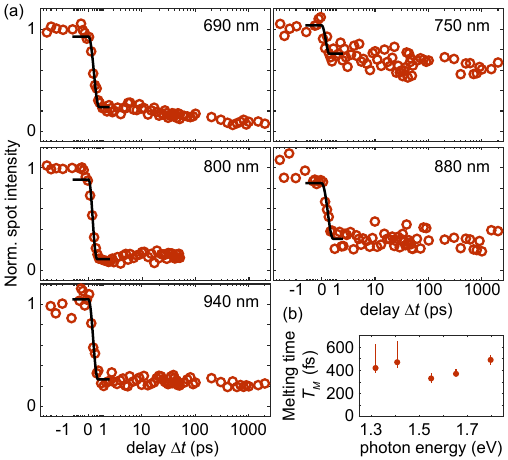}
\caption{
CDW dynamics at different excitation wavelengths in the visible spectrum.
(a) Irrespective of the incident pump photon energy, the second order superstructure reflections of the $(3\times3)$ phase (red circles) are suppressed within a half-cycle of approximately \SI{500}{\fs}. The black curves represent the results of the cosinusoidal fits.
(b) PLD melting time derived from the \SI{85}{\percent} intensity suppression as described in the text. We observe no significant influence of the pump photon energy on the temporal evolution of the dynamics.
}
\label{fig:WavelengthScan}
\end{figure}

The intensity of the $(3\times3)$ diffraction spots is a precise measure of the order parameter for the $(3\times3)$ to $(3\times1)$ transition, as it is intrinsically linked to the PLD amplitude. A precise determination of the relationship between the two quantities, however, must rely on dynamical diffraction simulations \cite{fujimoto1959} [c.f. Fig.~1(c) in the main text, and ref.~\cite{domrose2023} for a more detailed description]. Considering the structural refinement described in ref.~\cite{sorgel2006}, and averaging over an equal distribution in both sample thickness (ranging from \SI{40}{\nm} to \SI{60}{\nm}) and sample orientation (\SI{0.2}{\degree} total width around the specified zone axis), these simulations reproduce the expected quadratic behaviour for small PLD amplitudes \cite{Overhauser1971}, and deviations for larger distortions.

In our experiments, we evaluate the brightest spots situated in the ZOLZ at every temporal delay of the dynamics, while additionally considering the phononic background in the images. For the latter, we subtract the average inelastic scattering at late temporal delays $\Delta t>\SI{5}{\ps}$ and at a wave vector with a comparable magnitude from the individual delay curve of every reflection, before calculating the mean temporal response to the optical excitation by averaging over all individual delay curves. This procedure results in the complete suppression of $(3\times3)$ intensity when pumping above the phase transition threshold. The corresponding average delay curves for every pump fluence and excitation wavelength are displayed in Fig.~\ref{fig:FluenceScan} (grey data points) and Fig.~\ref{fig:WavelengthScan}, respectively. We derive the bare temporal response [red data points in Fig.~\ref{fig:FluenceScan}] of the PLD from a deconvolution of these curves with the instrument response function (green) measured directly before recording every individual delay scan (see also below in the corresponding section).

As described in the main text, the time constant associated with the dynamics of the $(3\times3)$ to $(3\times1)$ phase transformation $\tau$ could typically be extracted by assuming a displacive excitation mechanism \cite{zeiger1992}, fitting to a squared cosinusoidal function $I(\Delta t)^2$ where
\begin{align}
    I\left(\Delta t\right) =
    \begin{cases}
    1 & \Delta t \le 0\\
    I_0\,\text{cos}\left(\frac{\pi\Delta t}{T_\text{M}}\right)+I_0, & 0<\Delta t<T_\text{M}\\
    \end{cases}
\end{align}
Therein, $I_0=\frac{1+I_{\text{min}}}{2}$ with $I_{\text{min}}$ being the minimum reminiscent spot intensity after the PLD melting time $T_\text{M}$.

We find that the fitting results [black curves in Fig.~\ref{fig:FluenceScan} and \ref{fig:WavelengthScan}(a)] systematically underestimate the dynamics directly after time zero, such that, instead, we define $T_M$ based on the temporal delay where the average PLD spot intensity first falls below a value that corresponds to $\SI{85}{\percent}$ of the overall intensity drop at the respective fluence. The error bars displayed in Fig.~2(d) in the main text are derived by estimating the noise level in terms of the intensity fluctuations recorded before time zero, translating into the according uncertainty in determining the temporal delay associated with the $\SI{85}{\percent}$ criterion.

\section{Temporal evolution of first-order PLD diffraction spot intensities}
\label{sec:firstorderfluence}

\begin{figure}[htbp]
\centering
\includegraphics[width=\columnwidth]{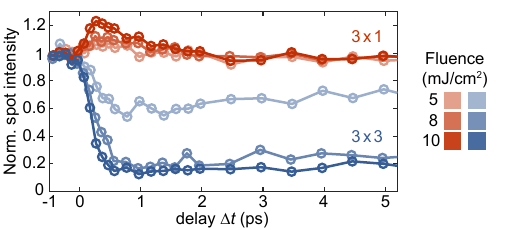}
\caption{
Fluence-dependent delay curves derived from first-order PLD diffraction spots.
Temporal evolution of intensities scattered into first-order diffraction spots of the $(3\times1)$ (red) and the $(3\times3)$ (blue) type for different excitation fluences. The high-temperature PLD spots display a transient overshoot that scales with the pump fluence, followed by a slower thermalization (see main text). In comparison, the $(3\times3)$ spots behave similar to their second-order counterparts.
}
\label{fig:First-order PLD}
\end{figure}

Both superperiodicities feature a three-fold stacking sequence \cite{sorgel2006}, such that first-order PLD spots require tilting the electron beam away from perpendicular incidence onto the individual material tri-layers. A representative electron diffractogram is depicted in Fig.~3(a) in the main text. The temporal response of these spots to the optical excitation is depicted in Fig.~\ref{fig:First-order PLD}. The first-order $(3\times3)$ spots (blue curves) are suppressed similarly to the second-order reflections, i.e., we observe a sub-picosecond suppression followed by a slower intensity recovery. In contrast, the $(3\times1)$ periodicities display an initial intensity increase of up to more than \SI{20}{\percent} during the melting of the low-temperature PLD, indicative of the redistribution of scattered intensity from the suppressed $(3\times3)$ to the $(3\times1)$ spots. The overshoot is followed by a slower, few-picosecond thermalization of the system and the corresponding intensity decay originating from the Debye-Waller effect \cite{Waldecker2017,Otto2021}.

\section{Electron pulse duration and instrument response function}
\label{sec:electronpulses}

\begin{figure}[htbp]
\centering
\includegraphics[width=\columnwidth]{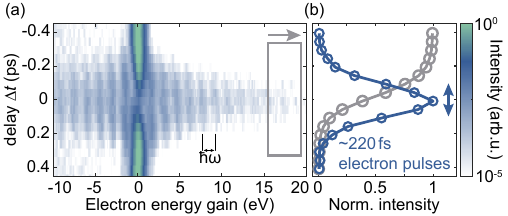}
\caption{
Independent characterization of the temporal resolution in the experiments.
(a) Laser-electron pulse cross-correlation, obtained from the inelastic interaction between the electron pulses with the laser-induced optical near-field, leading to the population of discrete side-bands in the electron spectra separated by the photon energy $\hbar\omega$.
(b) Instrument response function (grey) obtained from the cross-correlation by integrating the scattered intensity in the outer energy sidebands [grey rectangle in (a)]. The electron pulse shape (blue) is derived from the numerical derivative of the response function, yielding a pulse duration of approximately \SI{220}{\fs} for this particular measurement.
}
\label{fig:Pulse duration}
\end{figure}

\begin{figure*}[t]
\centering
\includegraphics[width=\textwidth]{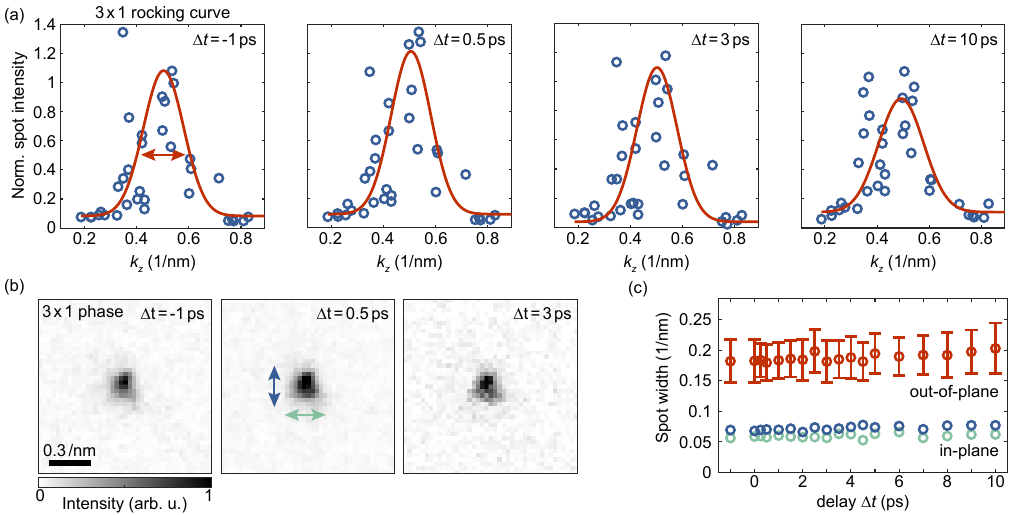}
\caption{
Three-dimensional temporal evolution of the $(3\times1)$ diffraction spot shape.
(a) $(3\times1)$ rocking curves as extracted from a single measurement under tilted sample illumination for an excitation fluence of \SI{7.5}{\mJ\per\cm^2}. At every temporal delay, every data point (blue circles) corresponds to the intensity scattered into an individual $(3\times1)$ diffraction spot situated at out-of-plane momentum $k_z$. The resulting rocking curve (red), centered around a value for $k_z$ that corresponds to the threefold periodicity along the material layer axis, preserves its width during the entire dynamics.
(b) In-plane spot profile of a representative individual $(3\times1)$ reflection. Similar to (a), the spot shape is independent on the temporal evolution of the phase transformation.
(c) Diffraction spot widths (FWHM) resulting from a fit of a gaussian spot shape to the data displayed in (a) and (b). The errorbars represent the \SI{68}{\%} confidence interval of the fit results for the respective in-plane (green and blue) and the out-of-plane values (red).
}
\label{fig:SpotShape}
\end{figure*}

The electron pulse duration is obtained by measuring the instantaneous inelastic interaction of the electron pulses with the optical near-field at the surface of the specimen (Photon-induced near-field electron microscopy, PINEM \cite{barwick2009,Feist2015a,bach2019,dahan2021}). Therein, individual electrons either gain or lose energy in multiple integers of the photon energy, leading to characteristic sidebands in the electron spectra when they are in spatiotemporal overlap with the laser illumination [cf. Fig.~\ref{fig:Pulse duration}(a)]. Scanning the temporal delay between the electron and the laser pulses in the sample plane, and summing up the intensity scattered into the outermost sideband at every temporal delay then gives a precise measure of the electron pulse duration [considering that the electron pulses are significantly longer than the laser pulses, Fig.~\ref{fig:Pulse duration}(b)]. Similarly, the temporal instrument response function, i.e., how the experimental setup would resolve a step-like temporal materials response to the optical excitation, can be obtained by a cumulative summation of the electron energy gain side, synonymous to an integration of the electron pulse shape along the temporal axis. The resulting instrument response functions are displayed in Fig.~2(b) in the main text and in Fig.~\ref{fig:FluenceScan}.

\section{Temporal evolution of the diffraction spot shape}
\label{sec:spotshape}

In the recorded diffractograms, the long-range coherence of the CDW is encoded in the temporal evolution of the shape of the individual CDW diffraction spots. As such, disorder along the in-plane direction would result in a broadening of the corresponding $(3\times1)$ reflections. In our experiments, such a broadening is absent regardless of the laser pump fluence, indicating that the phase transformation does not involve the formation of topological defects [c.f. Fig.~\ref{fig:SpotShape}(b) and (c)]. Similarly, the out-of-plane coherence, i.e., the CDW stacking sequence, can be inferred from a reconstruction of the CDW rocking curve [Fig.~\ref{fig:SpotShape}(a)]. To this end, electron diffractograms recorded under tilted-beam conditions feature multiple Laue zones [c.f. Fig.~3(a)] that allow to densely sample the CDW diffraction spot shape at different out-of-plane momenta \cite{domrose2023}. For the image series displayed in Fig.~\ref{fig:SpotShape}, we assume that the most intense $(3\times1)$ spots in our images are situated in the first-order Laue zone at an out-of-plane momentum corresponding to the threefold CDW stacking sequence. The resulting rocking curves are shown in Fig.~\ref{fig:SpotShape}(a). Generally, the occurrence of possible disorder along the layer direction is more difficult to resolve than comparable in-plane processes due to the strong influence of dynamical scattering on diffracted intensities. Nevertheless, a dimensional crossover during the dynamics would result in a pronounced intensity increase that is distributed homogeneously along the out-of-plane momentum $k_z$ \cite{domrose2023}. As such a dynamics is absent in our data, we conclude that the optical excitation does not induce transient disorder during the dynamics. Instead, the observed transient intensity overshoot and the subsequent $(3\times1)$ suppression described in the main text maintains the three-dimensional shape of the diffraction spots throughout the entire dynamics.

\newpage

\bibliography{TaTe2}


\end{document}